\documentclass[preprint,preprintnumbers,amsmath,amssymb,aps]{revtex4}

\usepackage{amssymb}
\usepackage{amsmath}
\usepackage{graphicx}
\usepackage{epsfig}
\usepackage{dcolumn}
\usepackage{bm}
\usepackage{fancyhdr}
\usepackage{epstopdf}

\def\be{\begin{align}}
\def\ee{\end{align}}
\def\bea{\begin{eqnarray}}
\def\eea{\end{eqnarray}}

\begin{document}

\title{On Quantum Coherence Effects in Photo and Solar Cells}
\author{Kimberly R. Chapin,$^1$ Konstantin Dorfman$^1$, Anatoly Svidzinsky$^{1,2}$ and Marlan O. Scully$^{1,2}$}
\affiliation{$^1$Texas A\&M University, College Station, TX 77843}
\affiliation{$^2$Princeton University, Princeton, NJ 08544}

\date{\today }

\begin{abstract}
We show that quantum coherence can increase the quantum efficiency 
of various thermodynamic systems. For example, we can enhance 
the quantum efficiency for a quantum dot photocell, a laser 
based solar cell and the photo-Carnot quantum heat engine. 
Our results are fully consistent with the laws of thermodynamics 
contrary to comments found in the paper of A.P. Kirk, 
Phys. Rev. Lett. {\bf 106}, 048703 (2011).
\end{abstract}

\pacs{05.70.-a, 88.40Nj,42.50Ar}
\maketitle

\section{Introduction}
In a recent article \cite{1}, quantum coherence was used to increase the quantum efficiency of a photocell. To put this in context, we recall that according to conventional wisdom, radiative recombination limits the quantum efficiency of a photocell; e.g., when illuminated by a monochromatic beam of solar photons (characterized by frequency $\nu_s$ and temperature $T_s$) this limit is $eV/\hbar\nu_s=\eta_c$ where $V$ is the maximum induced voltage, $\eta_c=1-T_a/T_s$ and $T_a$ is the ambient temperature, that is
\begin{equation}
eV=\hbar \nu_s\left(1-\frac{T_a}{T_s}\right)~.
\end{equation}

The physical picture we have in mind is shown in Fig. 1 of Ref \cite{1}, which adapted for the presented purposes, is enclosed here as Fig. 1a. The quantum dots in Fig. 1a form an ideal two level system and as explained in Ref \cite{1}, Eq (1) follows.

\begin{figure}[h]
\begin{center}
\epsfig{figure=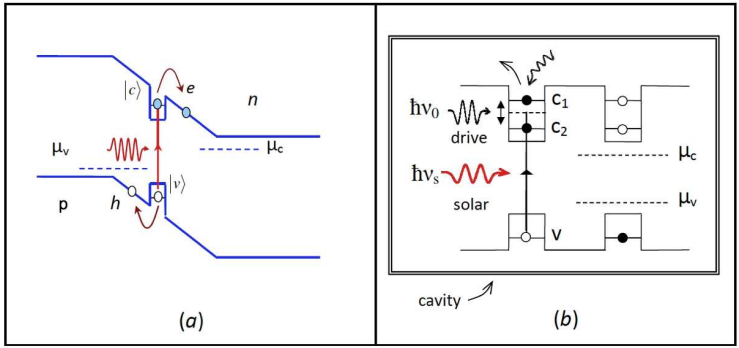, angle=270, width=10cm}
\end{center}
\caption{$(a)$ $p-n$ junction with quantum dot in the depletion region. Chemical potentials $\mu_v$ and $\mu_c$ are indicated by dashed lines. The ''built-in" field in the depletion layer separates electrons and holes and related to the voltage by $eV=\mu_c-\mu_v$, however electrons can radiatively recombine before being separated. $(b)$ Quantum dots having upper level conduction band states $|c_1\rangle$ and $|c_2\rangle$ are coherently driven by a field such that $\hbar\nu_{0}=\frac{1}{2}(\epsilon_{c_1}-\epsilon_{c_2})$. The monochromatic solar photons having energy $\hbar\nu_s$ are tuned to the mid point between the upper levels.}
\label{Fig:radiative} %
\end{figure}

A stimulus for Ref \cite{1} was the excellent review of solar cell physics \cite{2} in which we find the statement
\begin{quote}
``That leaves radiative recombination as the major [energy loss] process. Can this be avoided? The answer is no. If a radiative upward transition to generate the excitation is allowed, its reversal, the radiative downward transition must be allowed as well."
\end{quote}
However in \cite{1}, it is shown that it is possible to break detailed balance as in the case of lasing without inversion yielding a photocell quantum efficiency  $eV/\hbar \nu_s =\eta_c+\delta \eta$ where $\delta \eta$ represents a model dependent increase in efficiency. For example, in the toy model of Fig. 1b, we find $\delta\eta=\hbar\nu_0/\hbar\nu_s$, which implies

\begin{equation}
eV=\hbar \nu_s\left(1-\frac{T_a}{T_s}\right)+\hbar\nu_0~.
\end{equation}

Thus we now have (in principle) an open circuit voltage which exceeds the voltage given by Eq. (1). In \cite{1}, we say
\begin{quote}
``The preceding coherent drive model illustrates the role of quantum coherence in a simple way. However, it is possible to generate coherence $\rho_{1,2}$ without the use of an external field. For example, quantum noise induced coherence via Fano coupling yields an essentially equivalent result."
\end{quote}

For example, in a forthcoming article, we show that the power delivered to the load (given by the current times the voltage) is greater for a three-level system with Fano coupling, as depicted in Fig. 2.

\begin{figure}[h] %
\includegraphics[width=10cm,angle=0]{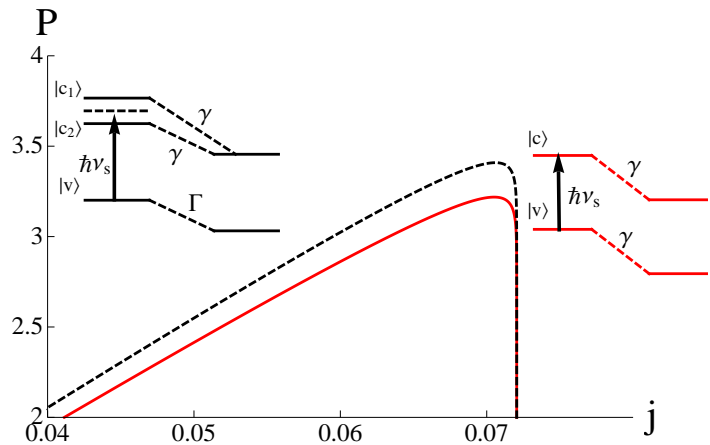} %
\caption{Power  generated by a solar cell  $P$ (arb. units) as a function of electric current $j$ (arb. units) through the cell for two-level system - (red solid line), and three-level system - (black dashed line).} %
\label{Fig:optical} %
\end{figure}

Clearly the discussion of \cite{1} based on our toy quantum dot photocell is concerned with issues of principal. Practical applications are not the point here. Nevertheless, A. Kirk attempts to investigate the limits of Ref. \cite{1} in a PRL article \cite{4} entitled ``Analysis of quantum coherent semiconductor quantum dot p-i-n junction photovoltaic cells." Although we want to encourage such studies, we disagree with many of his conclusions. For example, the following (incorrect) Kirk comments (K(i) i=1,2,3) and Scully replies S(i) are cases in point.

K(1): Scully argues that he has identified a PV cell that can generate more power than is emitted by the sun. 

S(1): This is Kirk's incorrect conclusion, not mine. The quantum efficiency determines the open circuit voltage, not the power.

K(2): As Harris shows in his 1989 work on lasing without inversion, Fano interference does not break detailed balance. 

S(2): I disagree, as does Harris, and I thank him for allowing me to report so.

K(3): The key component of the photo-Carnot engine is the new working fluid "phaseonium" . . .This renders the implied connection with increased solar PV efficiency void. 

S(3): This statement is void. Photons are the working fluid in the photo-Carnot engine, not phaseonium.

Further comments on Ref. \cite{4} will be addressed elsewhere. However, it is perhaps more constructive to first consider another simple laser based solar energy converter, as is given in the reply to Kirk \cite{5}, and further developed in the next section. Section III is devoted to a short discussion of the photo-Carnot quantum heat engine and concluding remarks are given in IV.

\section{The Laser is a Quantum Heat Engine and LWI enhances the quantum efficiency}
\begin{figure}[h]
\begin{center}
\epsfig{figure=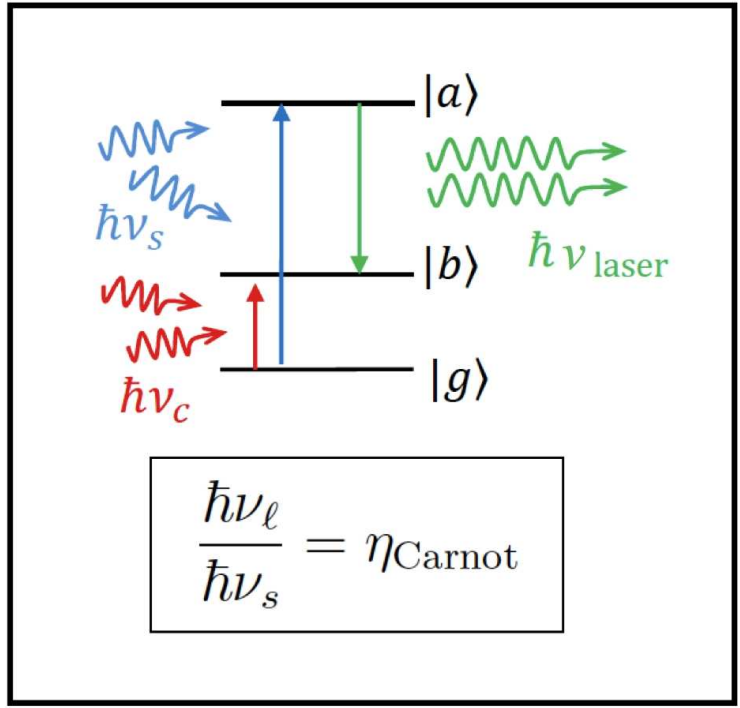, angle=270, width=10cm}
\end{center}
\caption{Thermal radiation from a ``hot" source (temperature $T_s$, frequency $\nu_s$) constitutes the energy reservoir while ``cold" (temperature $T_c$, frequency $\nu_c$) forms the entropy sink.  The combination produces coherent laser radiation, i.e., useful work with a quantum efficiency given by $\eta_{\text{Carnot}} = 1-T_c/T_h$.} %
\label{fig3}
\end{figure}

Quantum mechanics is the crowning achievement of twentieth century physics and continues to yield new fruit in the twenty-first century. For example, quantum coherence effects such as lasing without inversion \cite{6,7,8}, the Photo-Carnot Quantum heat engine \cite{9}, and the quantum photocell \cite{1} are topics of current research interest which are yielding new insights into thermodynamics and optics. No better way to explain this than to begin with the fact that the laser is a heat engine with Carnot quantum efficiency as depicted in Fig. \ref{fig3}. That is

\begin{equation}
\frac{\hbar\nu_{\ell}}{\hbar\nu_s}=\eta_{\text{Carnot}}=1-\frac{T_c}{T_h}~.
\end{equation}

\begin{figure}[h]
\begin{center}
\epsfig{figure=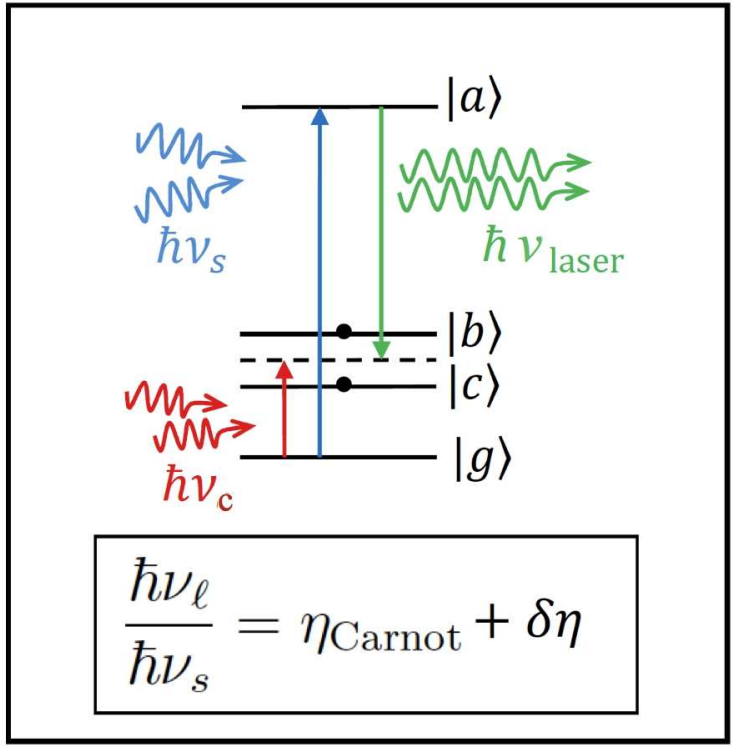, angle=270, width=10cm}
\end{center}
\caption{The atom of Fig. 3 is now replaced by an atom having a pair of closely spaced levels $b,c$. This double acts as the lower laser state when the laser is tuned midway between $b$ and $c$. Then the laser operates with a quantum efficiency $\eta_{\text{Carnot}}$ just as in the case of Fig. 3. However, when there is a bit of coherence between the levels $b$ and $c$ an extra quantum efficiency $\delta\eta$ is obtained. } %
\label{fig4}
\end{figure}

The plot thickens when we bring quantum coherence into the game. As indicated in Fig \ref{fig4}, if the lower level doublet has a small amount of coherence such that $|\rho_{bc}|<<\sqrt{\rho_{bb}\rho_{cc}}$ the quantum efficiency is now increased and is given by
\begin{equation}
\frac{\hbar\nu_{\ell}}{\hbar\nu_s}=\eta_{\text{Carnot}}+\delta\eta~,
\label{eq1}
\end{equation}
where $\delta\eta=kT_c|\rho_{bc}|/\hbar\nu_s\rho_{aa}$. The details of calculation are present in Appendix A.

These studies are the basis for lasing without inversion (LWI) and is further developed in the next paragraph on a laser solar cell enhanced by LWI and in the appendix. Furthermore, we show how this model allows coherent control of photo-Carnot engines.

\begin{figure}[h]
\begin{center}
\epsfig{figure=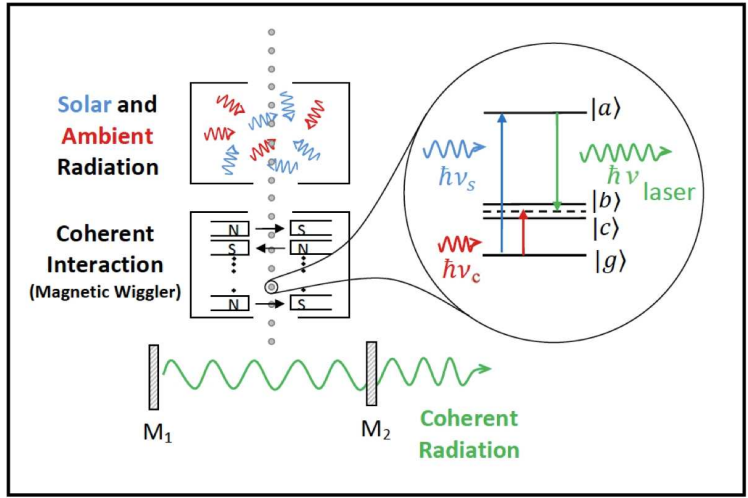, angle=270, width=10cm}
\end{center}
\caption{Atoms pass through a beam of ``hot" solar photons while interacting with ambient ``cold" light at the same time. The population of upper level $|a\rangle$ and the lower level doublet $|b\rangle, |c\rangle$ is governed by  Boltzmann factors characterized by temperatures $T_h$ and $T_c$ respectively. Then the atoms pass through a region where coherence $\rho_{bc}$ is generated via, e.g., a wiggler array, etc. Finally the atoms enter into the laser cavity where coherent laser light (tuned to the midpoint between $|b\rangle$ and $|c\rangle$) is generated.} %
\label{fig5}
\end{figure}

Let us consider the solar pumped laser of Fig. \ref{fig5}. As discussed previously, a thermally pumped laser is a quantum heat engine. An incoherent thermal pump serves as the energy source at $T_h$ that populates the upper laser level. The lower laser level doublet has small amount of coherence $|\rho_{bc}|<<\sqrt{\rho_{bb}\rho_{cc}}$ and is coupled to the ground state by ``cold" light at $T_c$ which serves as an entropy sink. The quantum efficiency is then given essentially by Eq. (\ref{eq1}). Thus, the quantum efficiency for the conversion of incoherent solar photons to coherent laser photons (useful work) is enhanced by the factor $\delta\eta$.

\section{The Photo-Carnot Engine}

	In \cite{8} we developed a quantum Carnot engine in which the atoms in the heat bath are given a small bit of quantum coherence. The induced quantum coherence becomes vanishingly small in the high-temperature limit at which we operate and the heat bath is essentially thermal. However, the phase $\phi$, associated with the atomic coherence, provides a new control parameter that can be varied to increase the temperature of the radiation field and to extract work from a single heat bath. The deep physics behind the second law of thermodynamics is not violated; nevertheless, the quantum Carnot engine has certain features that are not possible in a classical engine.
	
	Specifically in \cite{8} we proposed and analyzed a new kind of quantum Carnot engine powered by a special quantum heat bath consisting of phase coherent atoms (a.k.a. phasonium), which allows us to extract work from a single thermal reservoir. In this heat engine, radiation pressure drives the piston. Thus the radiation is the working fluid (analogous to steam), which is heated by a beam of hot atoms (analogous to coal) (Fig. 6).

\begin{figure}[h]
\begin{center}
\epsfig{figure=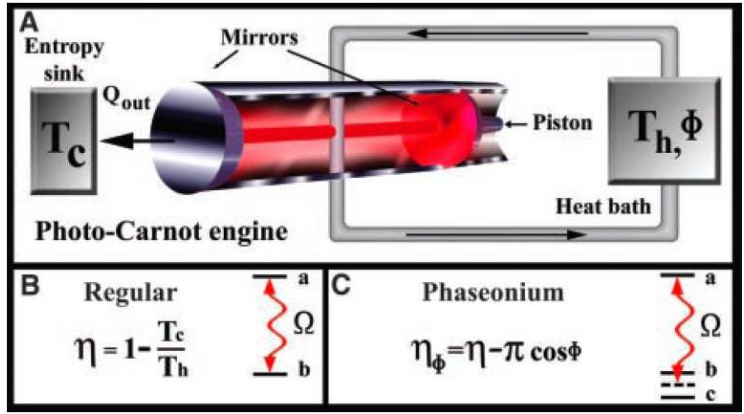, angle=270, width=10cm}
\end{center}
\caption{(A) Photo-Carnot engine in which radiation pressure from a thermally excited single-mode field drives a piston. Atoms flow through the engine and keep the field at a constant temperature $T_{rad}$ for the isothermal $1\rightarrow 2$ portion of the Carnot cycle. Upon exiting the engine, the bath atoms are cooler than when they entered and are reheated by interactions with the hohlraum at $T_h$ and "stored" in preparation for the next cycle. The combination of reheating and storing is depicted in (A) as the heat reservoir. A cold reservoir at $T_c$ provides the entropy sink. (B) Two-level atoms in a regular thermal distribution, determined by temperature $T_h$, heat the driving radiation to $T_{rad} = T_h$ such that the regular operating efficiency is given by $\eta$. (C) When the field is heated, however, by a phaseonium in which the ground state doublet has a small amount of coherence and the populations of levels $a$, $b$, and $c$, are thermally distributed, the field temperature is $T_{rad}>T_h$, and the operating efficiency is given by $\eta_{\phi}=\eta-\pi\cos\Phi$.}
\label{fig7}
\end{figure}

	Here, we can get work from a single bath by using quantum coherence. This is possible because quantum coherence allows us to break detailed balance between emission and absorption as in the case of LWI. There is a connection with Maxwell's demon. With phaseonium fuel, we get a kind of "sorting action" in which hot atoms emit photons, but cold atoms absorb less than they ordinarily would. Phaseonium fuel can use quantum coherence and interference to achieve single bath operation in the spirit of Maxwell's demon.

	Naturally, we do not claim to have a "perpetual mobile of the second kind." We do claim to be able to extract work from a single heat bath. It takes energy, e.g., from an external source of microwaves, to prepare the coherence, but we may view this energy as part of the refining process yielding "superoctane" quantum fuel. Surely the "price at the pump" of phaseonium is higher than regular fuel, but once the tank is full, a little quantum coherence allows our Quantum Heat Engine (QHE) to extract energy from the high-temperature heat bath more efficiently, to run faster, or to do both.

	Alternatively, we could incorporate the microwave generator (which produces the coherence) into the photo-Carnot engine as a kind of "quantum supercharger." The practicality and utility of the photo-Carnot engine are not the issue here. The fact is quantum control via quantum coherence provides a new tool in the study of thermodynamics.

In particular, the LWI form of quantum coherence is found to provide a demonesque control parameter that allows work to be extracted from a single heat bath. The total system entropy is constantly increasing, and the physics behind the second law is not violated. However, quantum coherence does allow certain features of quantum engine operation which are not possible with a classical heat engine.

\section{Conclusions}

In present work we demonstrated that quantum coherence can increase 
the useful work extracted from the laser based solar cell 
in particular and quantum heat engines in general. For example, 
as shown in \cite{1}, quantum coherence can yield (in principle) an 
increase in the quantum efficiency of the photocell which governs the 
open circuit voltage. Furthermore, the analysis of the laser-based 
solar cell enhanced by LWI shows similar enhancement of 
quantum efficiency of the heat engine due to quantum coherence. 
Finally in the photocell with quantum coherence generated by 
quantum noise via Fano interference we can obtain enhanced
generated power.

\acknowledgements{This work was stimulated by a workshop
financed by the Texas Engineering Experiment Station. Support
from The Welch Foundation (A-1261), The Office of Naval Research, and The King Abdulaziz City of Science and Technology (KACST) is gratefully acknowledged. }

\section*{Appendix A}

\renewcommand{\theequation}{A\arabic{equation}}
\renewcommand{\thefigure}{A\arabic{figure}}
   \setcounter{equation}{0}  
   \setcounter{figure}{0}



In the scheme of Fig. 3, the quantum efficiency of the laser, $\eta_{\ell}$, is given by $\hbar\nu_{\ell}/\hbar\nu_s$. Following the convention established by  Scovil and Schulz-DuBois \cite{11} which treats the three level maser as a quantum heat engine. The  Boltzmann factors involved in our system are:

\begin{equation}\label{eq:A1}
\frac{n_b}{n_a}=\frac{n_b}{n_g}\cdot\frac{n_g}{n_a}=e^{-\hbar\nu_c/kT_c}\cdot e^{\hbar\nu_s/kT_h}
\end{equation}

The threshold condition for lasing $n_b=n_a$ then implies $\hbar\nu_c/kT_c=\hbar\nu_s/kT_h$ and using $\hbar\nu_s-\hbar\nu_{\ell}=\hbar\nu_c$ we find
\begin{equation}
\frac{\hbar\nu_{\ell}}{\hbar\nu_s}=1-\frac{T_c}{T_s}
\end{equation}

Let us next consider the LWI-like system of Fig. 4 or Fig. 5 in which the lower level doublet has a small amount of coherence such that $|\rho_{bc}|<<\sqrt{\rho_{bb}\rho_{cc}}$. As is shown in detail in chapter 7 of Ref. \cite{8}, the laser field ${\cal{E}}$ is governed by
\begin{equation}\label{eq:A3}
\dot{\cal{E}}=\kappa (2\rho_{aa}-\rho_{bb}-\rho_{cc}-\rho_{bc}-\rho_{cb})\cal{E}
\end{equation}
where $\kappa$ is an overall constant. To a good approximation, the populations are determined by $T_h$ and $T_c$.

At threshold, Eq (\ref{eq:A3}) implies that
\begin{equation}
1-\frac{\rho_{bb}}{\rho_{aa}}-\frac{(\rho_{bc}+\text{c.c.})}{2\rho_{aa}}=0
\end{equation}
where we have used the fact that $\rho_{bb}\cong \rho_{cc}$. Noting that

\begin{equation}
\frac{\rho_{bb}}{\rho_{aa}}=e^{-[\hbar\nu_c/kT_c-\hbar\nu_s/kT_h]}~,
\end{equation}
 the threshold relation (A4) then yields
\begin{equation}
\frac{\hbar\nu_s}{kT_h}-\frac{\hbar (\nu_s-\nu_{\ell})}{kT_c}=ln\left[1-\frac{(\rho_{bc}+\text{c.c.})}{2\rho_{aa}}\right]~.
\end{equation}
Finally we take $\rho_{bc}=|\rho_{bc}|e^{i\pi}$ and for weak coherence $|\rho_{bc}|\ll\rho_{aa}$ obtain  
\begin{equation}
\frac{\hbar\nu_{\ell}}{\hbar\nu_s}=\eta_{\text{Carnot}}+\delta\eta~,
\end{equation}
where $\delta\eta=kT_c|\rho_{bc}|/\hbar\nu_s\rho_{aa}$.

\end{document}